\documentclass[trackchanges,twocolumn]{aastex7}
\makeatletter
\let\frontmatter@title@above=\relax
\usepackage{graphicx}
\usepackage{multirow}
\usepackage{txfonts}
\usepackage{chngcntr}
\usepackage{natbib}
\usepackage{url}
\usepackage{hyperref}
\usepackage{longtable}
\usepackage{tabularx}
\usepackage{lipsum}
\usepackage{array}
\usepackage{rotating}
\usepackage{changepage}
\usepackage{epstopdf}

\shorttitle{X-ray Properties of MAXI J1744-294}
\shortauthors{Chatterjee, Mondal, Palit, et al.}

\begin{document}

\title{Distinct Fe-K Line Complexes in MAXI J1744-294 Revealed by XRISM High-Resolution Spectroscopy}

\author[0000-0002-6252-3750]{Kaushik Chatterjee}
\affiliation{South-Western Institute for Astronomy Research (SWIFAR), Yunnan University, Kunming, Yunnan 650500, People's Republic of China}
\affiliation{Key Laboratory of Survey Science of Yunnan Province, Yunnan University, Kunming, Yunnan 650500, People's Republic of China}
\email[show]{mails.kc.physics@gmail.com, kaushik@ynu.edu.cn (KC)}

\author[0000-0003-0793-6066]{Santanu Mondal}
\affiliation{Indian Institute of Astrophysics, II Block, Koramangala, Bengaluru 560034, Karnataka, India}
\email[show]{\\santanuicsp@gmail.com (SM)}

\author[0000-0002-4533-3170]{Biswaraj Palit}
\affiliation{Nicolaus Copernicus Astronomical Center, Polish Academy of Sciences, ul. Bartycka 18, 00-716 Warsaw, Poland}
\email[show]{\\bpalit@camk.edu.pl (BP)}

\author[0000-0002-7782-5719]{Chandra B. Singh}
\affiliation{South-Western Institute for Astronomy Research (SWIFAR), Yunnan University, Kunming, Yunnan 650500, People's Republic of China}
\affiliation{Key Laboratory of Survey Science of Yunnan Province, Yunnan University, Kunming, Yunnan 650500, People's Republic of China}
\email[]{chandrasingh@ynu.edu.cn}

\author[0000-0002-6640-0301]{Sujoy Kumar Nath}
\affiliation{Indian Center for Space Physics,  466 Barakhola, Netai Nagar, Kolkata 700099, India}
\email[]{sujoynath0007@gmail.com}

\author[0000-0002-5900-9785]{Mayukh Pahari}
\affiliation{Department of Physics, Indian Institute of Technology Hyderabad, Hyderabad, Kandi, 502285 Sangareddy, India}
\email[]{mayukh@phy.iith.ac.in}

\author[0000-0001-7225-2475]{Brajesh Kumar}
\affiliation{South-Western Institute for Astronomy Research (SWIFAR), Yunnan University, Kunming, Yunnan 650500, People's Republic of China}
\affiliation{Key Laboratory of Survey Science of Yunnan Province, Yunnan University, Kunming, Yunnan 650500, People's Republic of China}
\email{brajesh@ynu.edu.cn, brajesharies@gmail.com}

\author[0000-0003-3901-8403]{Wei Wang}
\affiliation{Department of Astronomy, School of Physics and Technology, Wuhan University, Wuhan, People's Republic of China}
\email[]{wangwei2017@whu.edu.cn}

\author[0000-0002-5617-3117]{Hsiang-Kuang Chang}
\affiliation{Institute of Astronomy, National Tsing Hua University, Hsinchu 300044, Taiwan, Republic of China}
\email[]{hkchang@mx.nthu.edu.tw}

\author[0000-0003-1295-2909]{Xiaowei Liu}
\affiliation{South-Western Institute for Astronomy Research (SWIFAR), Yunnan University, Kunming, Yunnan 650500, People's Republic of China}
\affiliation{Key Laboratory of Survey Science of Yunnan Province, Yunnan University, Kunming, Yunnan 650500, People's Republic of China}
\email[]{x.liu@ynu.edu.cn}


\begin{abstract}

The newly discovered Galactic transient MAXI J1744-294 went into its first X-ray outburst in 2025. We study the spectral properties of this source in 
the $2-10$~keV energy band during this outburst using X-ray data from the XRISM satellite for both of its Resolve and Xtend instruments, taken on March 03, 
2025. High-resolution spectroscopy has revealed, for the first time, complex iron line features in this source, corresponding to distinct components of 
Fe XXV emission and Fe XXVI absorption lines. Such a detailed structure has not been reported in other low-mass X-ray binaries to date, prior to the XRISM 
era. Our analysis shows that the line complexes arise from two highly ionized plasmas with ionization rate $\sim 10^3$ erg cm s$^{-1}$ with distinct turbulent
velocities—one broad (v$_{\rm turb} \approx$ 2513 km s$^{-1}$) from hot gas at the inner accretion disk and one narrow (v$_{\rm turb} \approx$ 153 km s$^{-1}$)
scattered by nearby photoionized gas. These results offer new insight into the reprocessing of continuum in stratified media, either in the accretion 
disk or winds, or both, for XRBs in the soft state. The data are well described by models with spin, mass of the black hole, and accretion disk inclination 
$0.63-0.70$, $7.9 \pm 2.2$~M$_\odot$, and $19-24^\circ$. The fitted spectral model parameters suggest that the source is in the soft spectral state. The 
source is situated in a crowded field near the Galactic center, resulting in a large hydrogen column density. 

\end{abstract}



\section{Introduction}

Transient compact objects show variability signatures in the spectral and temporal properties during their outburst phases. These properties are strongly 
correlated with each other across different spectral states \citep[][and references therein]{2006ARA&A..44...49R, ChakrabartiEtal2008A&A...489L..41C,
2016ApJS..222...15T}. In different spectral states, the appearance and disappearance of timing phenomena like quasi-periodic oscillations, or QPOs, have 
been observed \citep[e.g.,][and references therein]{2002ApJ...572..392B,2005ApJ...629..403C, Mottaetal2011MNRAS.418.2292M,2014ApJ...786....4M,
2020MNRAS.493.2452C, 2024ApJ...977..148C, 2025ApJ...987...44C}. Additionally, jets or outflows are often observed phenomena that leave imprints in different 
spectral states \citep{1999A&A...351..185C,2000ApJ...543..373D, 1980A&A....86..121S, 2002ApJ...573L..35C,2021ApJ...920...41M}. Therefore, their X-ray 
spectra can tell us the underlying physical processes responsible for the spectral features. 

The observed spectral features mainly come from two distinct regions: one is the inner hot puffed-up region, or so-called corona, and the second 
component is the cold Keplerian disk. The seed photons from the cold disk produce the blackbody spectrum, and some fraction of these photons is 
upscattered by the corona and gives rise to the hard X-ray radiation \citep[][and references therein]{1980A&A....86..121S, 1993ApJ...413..507H, 
1995ApJ...455..623C, 2007A&ARv..15....1D}. Some of the hard photons from the corona fall back to the disk and come out as the so-called reflection 
component \citep{1991MNRAS.249..352G, 2005MNRAS.358..211R} leaving a hump above 10 keV \citep{2005MNRAS.358..211R, 2010ApJ...718..695G}, which is 
believed to be the origin of the Fe K fluorescent line around 6.4 keV \citep{1989MNRAS.238..729F, 1983ASPRv...2..189P}. However, the origin 
of the hump above 10 keV can also be from the radiation coming from the bulk motion Comptonization effect in the jet 
\citep{TitarchukShrader2005,2021ApJ...920...41M}. 
All these components of the radiation are informative in understanding the accretion behavior near the BH due to the strong gravitational 
effect, reflected in the Fe-K$\alpha$ line profile \citep{2006ApJ...652.1028B, 2014ApJ...787...83M}. The line can break into a double horn if the 
accretion disk moves much closer to the BH \citep[][and references therein]{1989MNRAS.238..729F, 1996MNRAS.282.1038I}, helping to estimate the spin 
of the central BH \citep[][and references therein]{2005ApJS..157..335L, 2024ApJ...975..257M}.

Apart from the Fe-K$\alpha$ line, there may be the presence of other Fe-line components also in the $6.3-7.3$~keV energy band. Multiple iron 
line emissions can be observed, which might be associated with the Galactic diffuse X-ray emission, or it can also be due to the physical conditions 
of the accreting matter \citep[see][for numerical simulations]{MondalEtal2021MNRAS.505.1071M}. The $\sim 6.6-6.7$~keV iron line emission plays a 
significant role in the context of our Galaxy. The emission line at $\sim 6.6-6.7$~keV can be identified as a combination of the $6.63$, $6.67$, and 
$6.70$~keV lines of Fe XXV ions, which could be produced through the recombination process in the accretion disk corona \citep{2000ApJS..131..571A}. 
The measurements of these features require high-quality X-ray data. The spectacular resolution of the XRISM \citep{TashiroXRISM2021SPIE11444E..22T} 
data can reveal these multiple line peaks and their origin.

The new Galactic source MAXI J1744-294 was reported on January 02, 2025 \citep{2025ATel16975....1K, 2025ATel16983....1N, 2025ATel17009....1W} by the 
Monitor of All-sky X-ray Image or, MAXI \citep{2009PASJ...61..999M} with a flux level of $\sim 250$ mCrab. From the follow-up observations of Swift/XRT, 
two sources were detected near Sgr A$^*$, one of which is consistent with the position of the neutron star low-mass X-ray binary AX J1745.6-2901, while 
the other one is an uncatalogued target at coordinates 17:45:41.93, -29:00:35 in J2000 coordinates \citep{2025ATel17010....1H}. Later on February 06, 
2025, {\it NuSTAR} also confirmed some activity from this new source \citep{2025ATel17063....1M}. It was observed on February 11-12, 2025, using {\it 
NICER} data by \citep{2025ATel17040....1J}. They reported the presence of an absorbed power-law along with a disk blackbody component in the $2-10$~keV 
energy band with an iron line at $6.6 \pm 0.9$~keV. The authors reported a high column density of $N_H = (11 \pm 1) \times 10^{22}$~cm$^{-2}$. This 
source is located very close to the Galactic center with (R.A., Dec.) = (266.116 deg, -29.433 deg). For the estimation of flux in the $2-10$ ~keV, 
\citet{2025ATel17063....1M} assumed a distance of $8$~kpc. Their estimated spectral parameters are consistent with those of a low-mass black hole X-ray 
binary in the soft state. Recently, \citet{SeshadriMajumderEtal2025arXiv250603774M} analyzed the {\it IXPE} data of the source and could not 
detect any polarization signature. Along the same line, \citet{2025arXiv250617050M} found only $\sim 1.3 \%$ polarization degree using IXPE data and 
constrained the disk inclination angle of the source in a broad range of $38-71^\circ$. A possible cause may be due to the presence of the soft spectral 
state, where a significantly low or null detection is favourable for several reasons \citep[see][]{2024ApJ...975..257M}. 

In this work, we have analyzed XRISM spectra for both the {\it Resolve} and {\it Xtend} instruments in the $2-10$~keV energy band of the newly detected 
Galactic source MAXI J1744-294 during its recent outburst in 2025. We have implemented some phenomenological and physical models for this analysis. 
In the next section, we describe the data reduction procedure. In \S3, we discuss the spectral analysis and results. Finally, we draw our conclusions in 
\S4.

\section{Data Reduction}

We have used XRISM data from both instruments that observed the source on March 03, 2025 (MJD 60737; Obs ID 901002010), for this work. 
The data reduction procedure is discussed below. 

The data of the new Galactic source MAXI J1744-294 is downloaded from the publicly available  
\href{https://heasarc.gsfc.nasa.gov/cgi-bin/W3Browse/w3browse.pl}{HEASARC archive}. Although level-2 cleaned event files are already present in
the downloaded data, we reprocessed the data using the {\tt xapipeline} task for both the Resolve and Xtend instruments separately. After the 
production of the level-2 cleaned files, as recommended in the 
\href{https://heasarc.gsfc.nasa.gov/docs/xrism/analysis/quickstart/xrism_quick_start_guide_v2p3_240918a.pdf}{Quick-Start Guide Version 2.3}, we 
used additional screening for pulse rise time, event type, and status. We also excluded the events from pixel number 27, as it is recommended for
calibration uncertainties. We use the {\tt XSELECT} task to extract spectra and light curve files for science analysis. Then we use {\tt rslmkrmf}, 
and {\tt xtdrmf} commands to make response files for the Resolve and Xtend instruments. We have used the extra-large (XL) type of response file for 
Resolve. Next, using the {\tt xaexpmap} command, we create a standard exposure map for both instruments. Then, we create ancillary response files 
for both instruments using the {\tt xaarfgen} command. We have also created the non-X-ray background for Resolve using the {\tt rslnxbgen} command.

\section{Spectral Analysis and Results}

We perform spectral analysis of the newly detected Galactic source MAXI J1744-294 using the processed data for science results. We used {\tt
\href{https://heasarc.gsfc.nasa.gov/xanadu/xspec/}{XSPEC}} \citep{1996ASPC..101...17A} to model the data using both phenomenological and physical 
\href{https://heasarc.gsfc.nasa.gov/xanadu/xspec/manual/node129.html}{models}. To account for the continuum emission, we have used combinations 
of models by using a {\tt \href{https://heasarc.gsfc.nasa.gov/xanadu/xspec/manual/node165.html}{disk blackbody}}, a {\tt 
\href{https://heasarc.gsfc.nasa.gov/xanadu/xspec/manual/node189.html}{kerrbb}}, and a {\tt 
\href{https://heasarc.gsfc.nasa.gov/xanadu/xspec/manual/node221.html}{power-law}} model. We have also used the Two-Component Advective 
Flow model \citep[{\tt TCAF};][]{1995ApJ...455..623C} including jet or {\tt JeTCAF} \citep{2021ApJ...920...41M} as part of the physical modeling. 
For the interstellar absorption, we have used the multiplicative absorption model component {\tt 
\href{https://heasarc.gsfc.nasa.gov/xanadu/xspec/manual/node279.html}{tbabs}}. Different combinations of these four models can fit the continuum 
satisfactorily. However, some line features are also present in the spectrum, particularly near the $6.3-7.3$~keV energy band. Thus, we added 
multiple {\tt \href{https://heasarc.gsfc.nasa.gov/xanadu/xspec/manual/node181.html}{Gaussian}} lines to fit these lines. This model combination 
is applied to both instruments. 

Furthermore, we have also performed photoionization modeling in the 5-8 keV energy band to study the distinct Iron line emission complexes 
and their origin. We discuss all these results in broad detail in the following subsections.

\subsection{Continuum Modeling}

\begin{figure}[!h]
\vskip 0.2cm
\centering
\vbox{\hskip -0.95cm
\includegraphics[width=8.3truecm,angle=0]{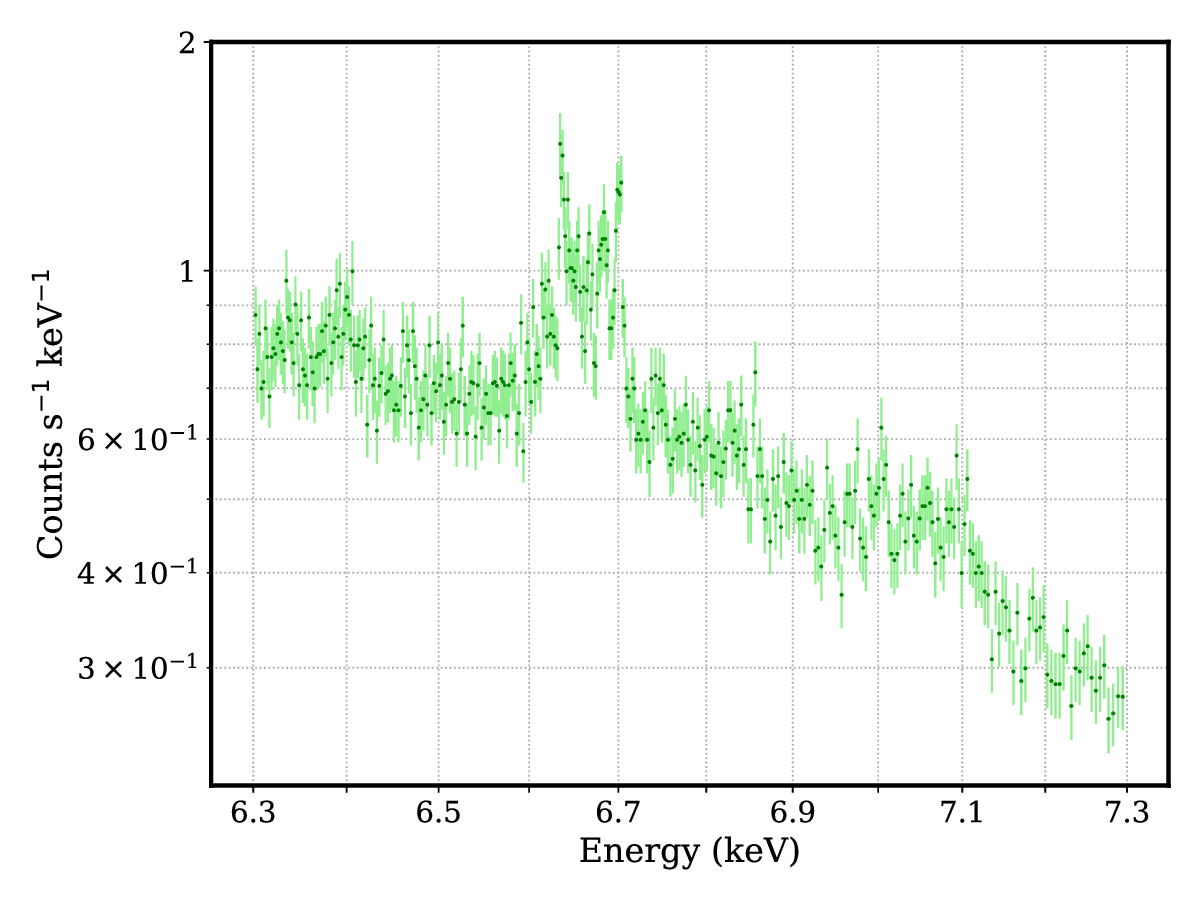}
\includegraphics[width=8.5truecm,angle=0]{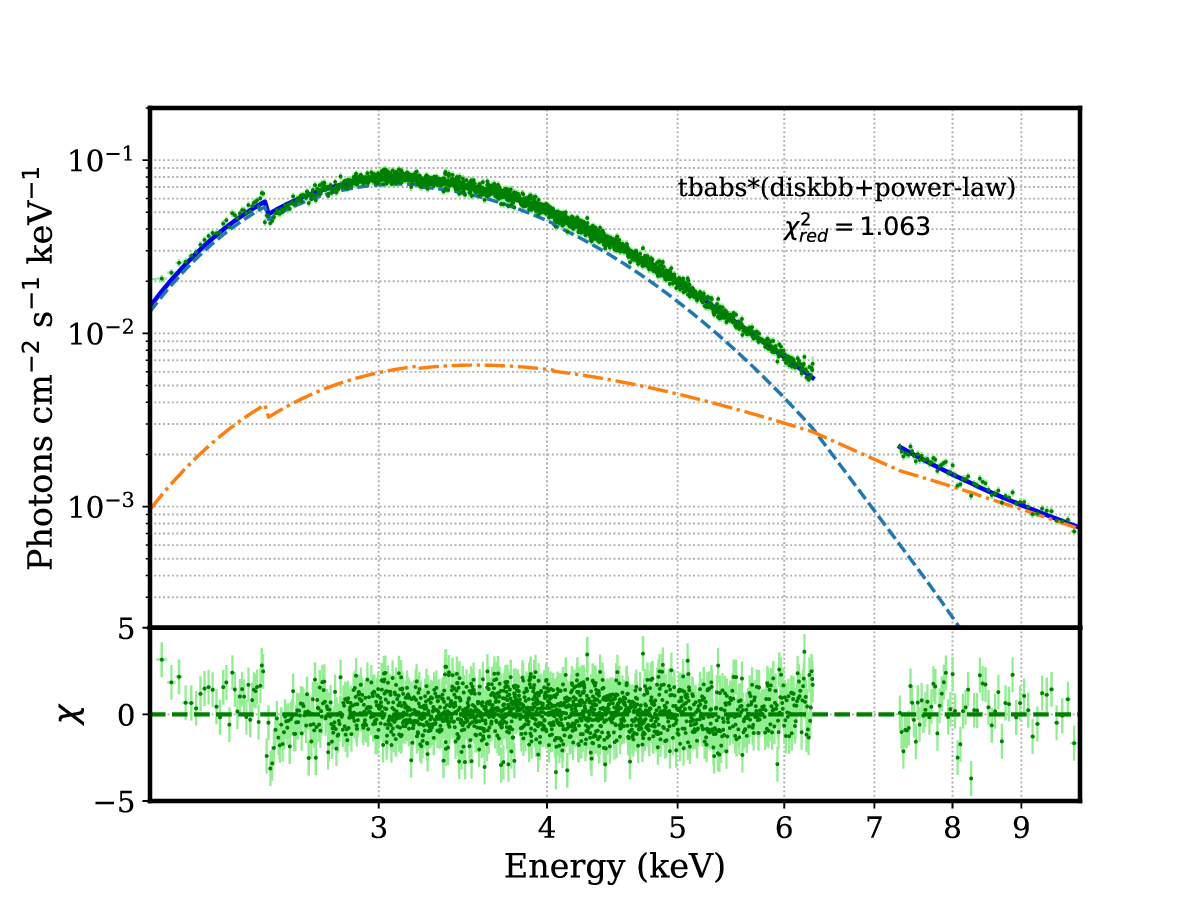}}
\caption{The Resolve spectrum in the range 6.3-7.3 keV is shown in the upper panel. The lower panel represents the fitted continuum in the $2-10$~keV 
energy band, ignoring the 6.3-7.3 keV region.} 
\label{fig:figa1}
\end{figure}

First, we looked at the spectra of Resolve in the $6.3-7.3$~keV. This is shown in the upper panel of \autoref{fig:figa1}. This shows that 
distinct multiple line features are present in the $6.3-7.3$ keV energy band. There are line emission profiles in the Resolve spectrum, which might be
related to the Fe-XXV line. That prompted us to first analyze the continuum spectrum and find a suitable model combination for the continuum emission. 
For that, we ignored the $6.3-7.3$~keV energy region and fitted the data using the model combination {\tt tbabs*(diskbb + po)}. We found that the model 
best fits the continuum with $\chi^2_{red} = 1.06$, as shown in the lower panel of \autoref{fig:figa1}.

\begin{figure}[!h]
\vskip 0.2cm
\centering
\vbox{\hskip -0.95cm
\includegraphics[width=8.3truecm,angle=0]{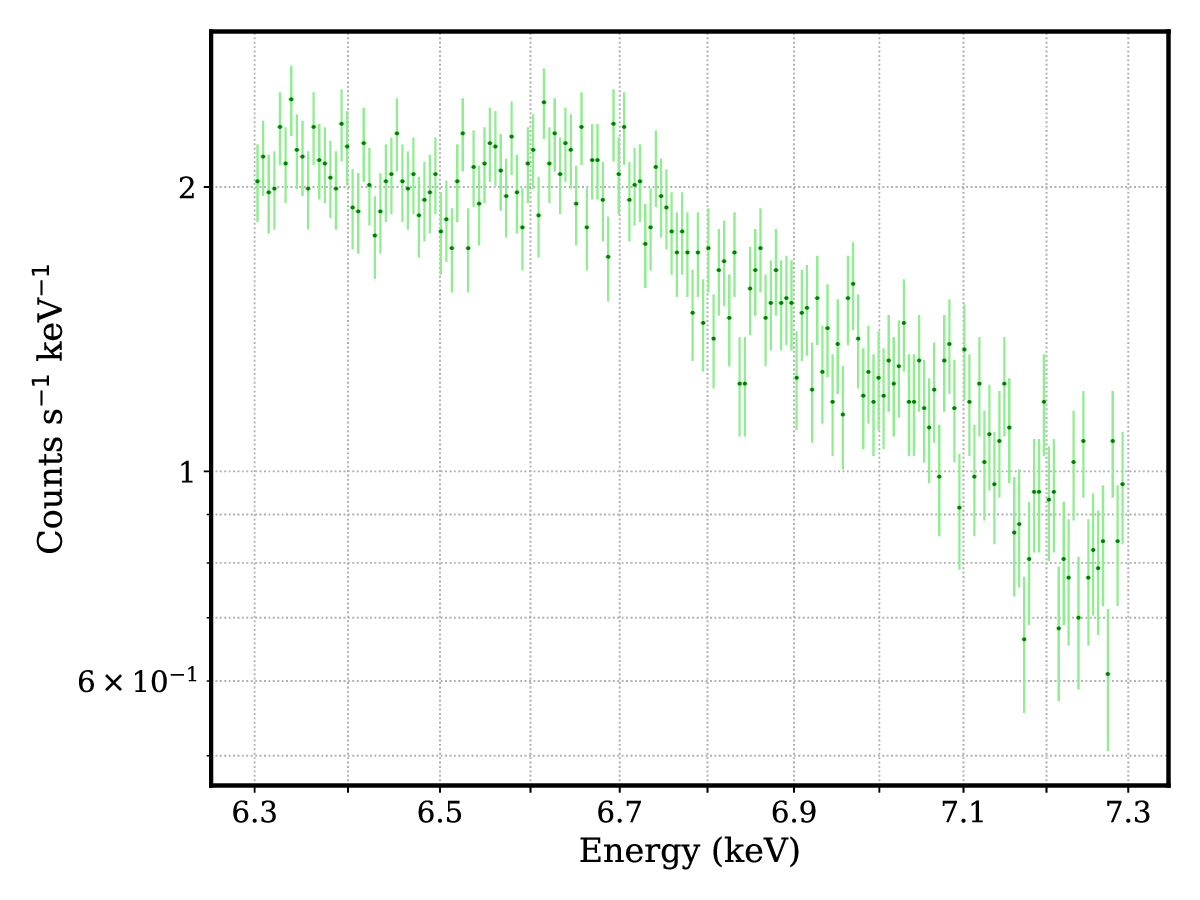}
\includegraphics[width=8.5truecm,angle=0]{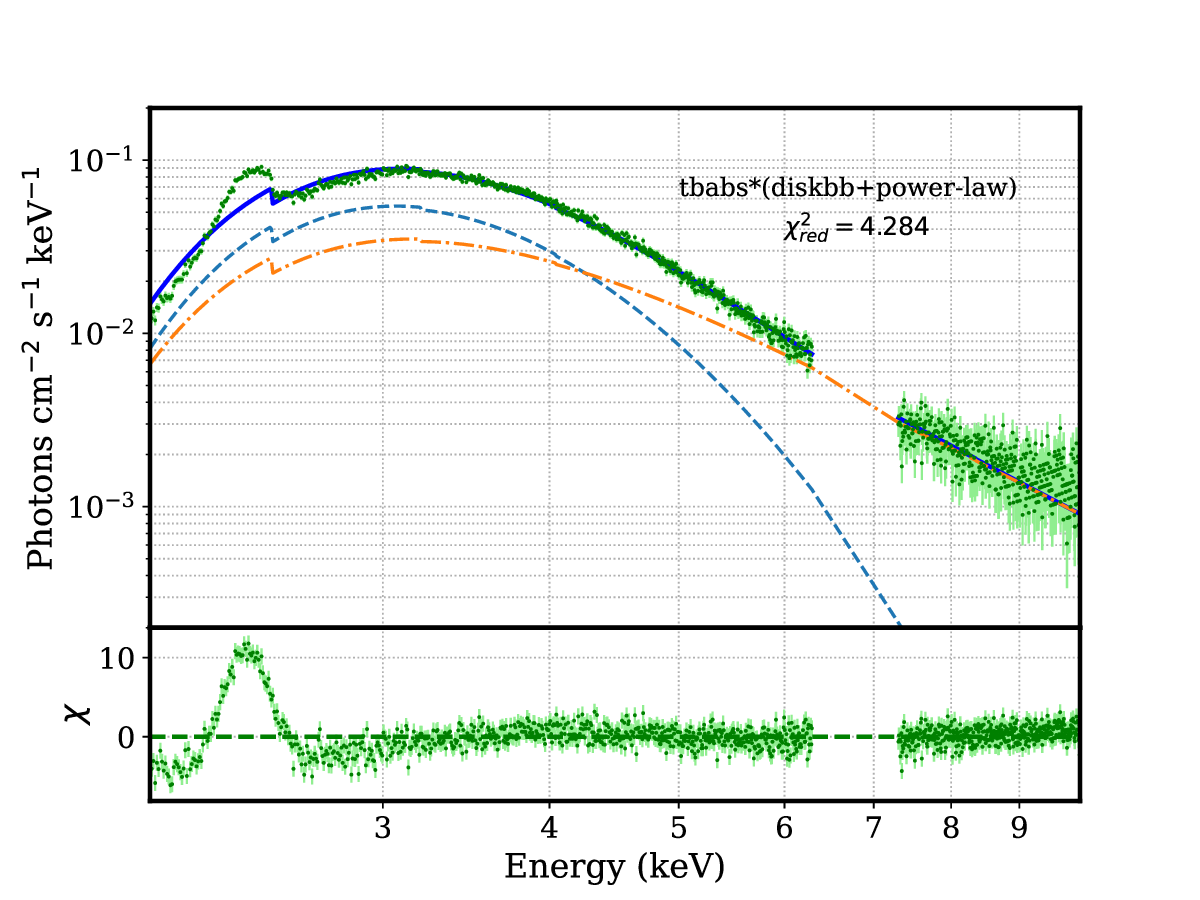}
\includegraphics[width=8.5truecm,angle=0]{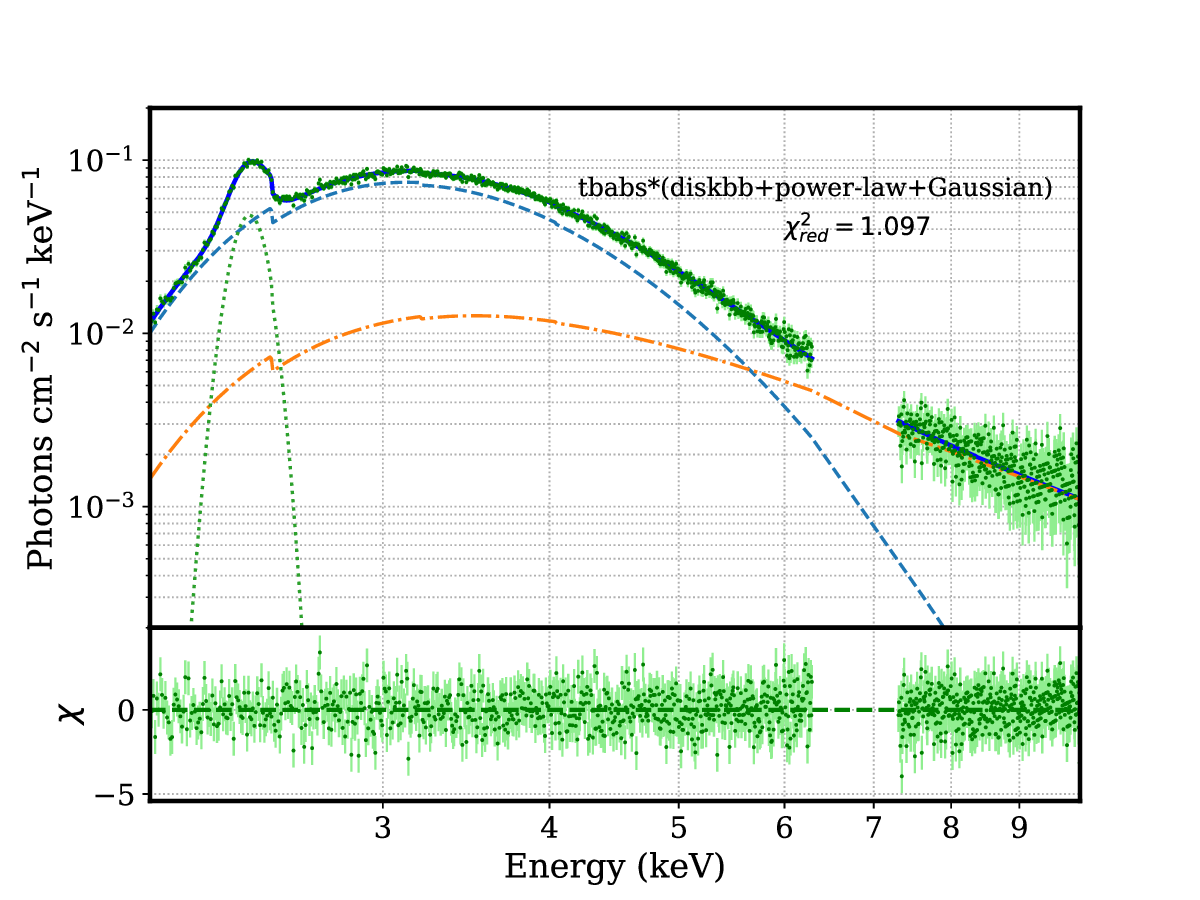}}
\caption{The Xtend spectrum in the range 6.3-7.3 keV is shown in the upper panel. Middle and lower panels represent the fitted continuum in the $2-0$
keV energy band, ignoring the $6.3-7.3$ keV region.}
\label{fig:figa2}
\end{figure}

Next, we looked at the Xtend spectrum and performed the same exercise. The upper panel of \autoref{fig:figa2} shows Fe-line features in the 
spectrum from Xtend in the $6.3-7.3$~keV energy band. Unlike the Resolve spectrum, this is not clearly resolved in the $6.3-7.3$~keV energy band, whether 
there are multiple line features or not. However, both spectra can certainly show the presence of Fe line profile in the energy range of $6.3-7.3$~keV.
Thus, we fit the continuum, excluding this energy band. For the Xtend spectrum, we found that the previous continuum model combination was not good 
enough. There is an excess of flux $\sim 2.4$~keV, which could be an instrumental calibration effect rather than astrophysical. The $\chi^2_{red} = 
4.284$. This can be noticed in the middle panel of \autoref{fig:figa2}. By adding another Gaussian ({\tt ga}) model, we achieve the best fit for the 
continuum. The model combination reads as {\tt tbabs*(diskbb + po + ga)} for which $\chi^2_{red} = 1.09$, as shown in the lower panel of \autoref{fig:figa2}.

The similar spectral features have been observed while fitted using other models in the following subsections. The details of the model fitting 
and parameters are described below.

\subsection{Continuum and Emission Line Modeling}

Here, we describe the results and model fitted parameters from the combined continuum and line emission models.

\subsubsection{{\tt diskbb} and {\tt power-law}}

\begin{figure}[!h]
\vskip 0.2cm
\centering
\vbox{
\includegraphics[width=8.5truecm,angle=0]{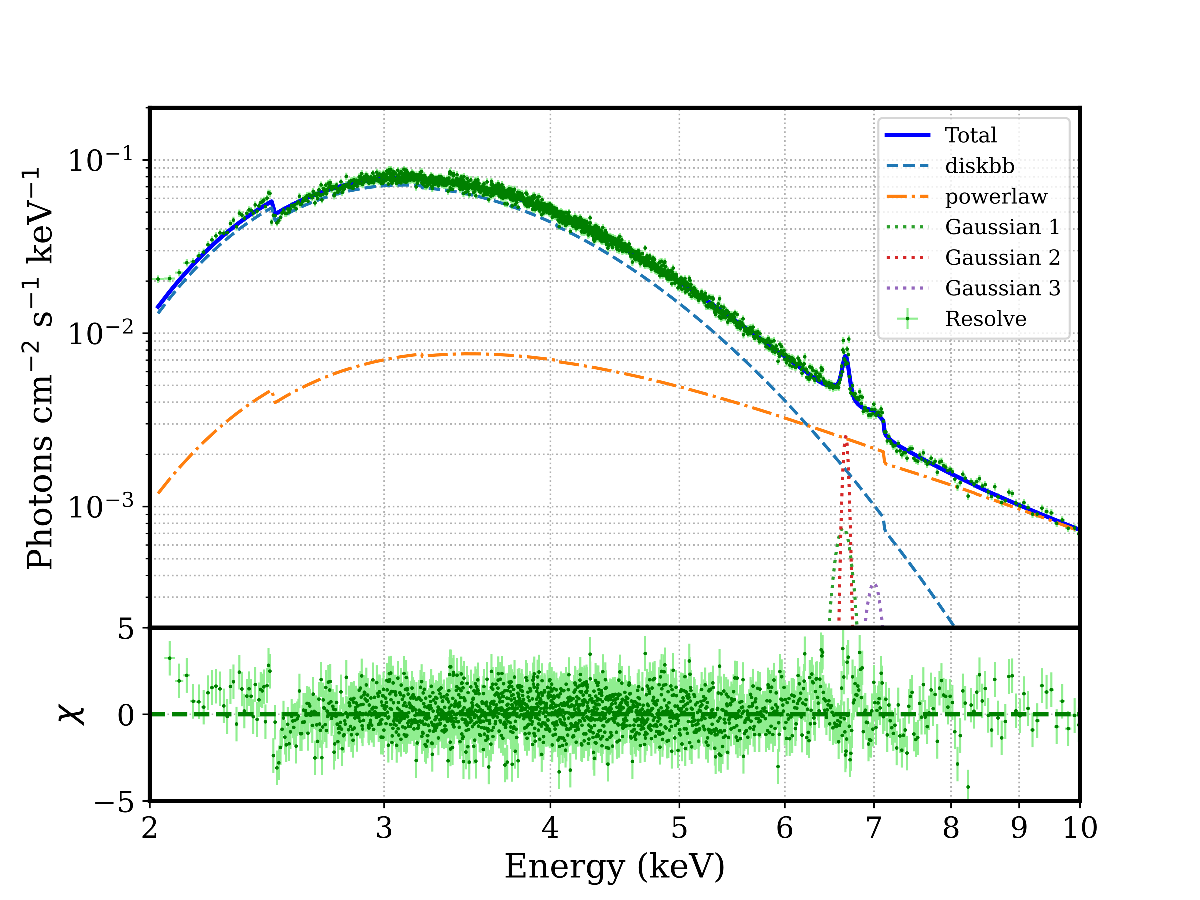}
\includegraphics[width=8.5truecm,angle=0]{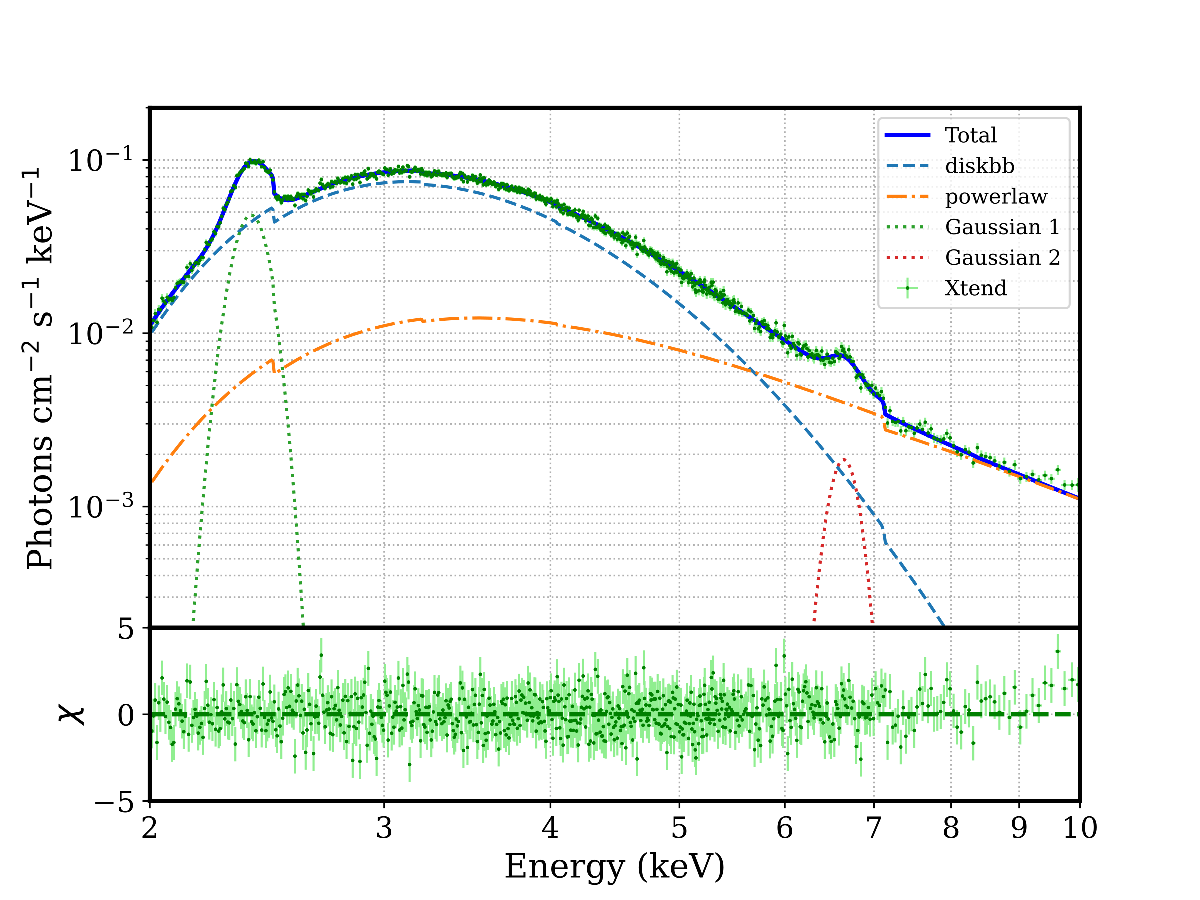}}
\caption{\label{fig:diskbb} Model-fitted unfolded XRISM spectra for Resolve (top panel) and Xtend (bottom panel) instruments, respectively, by
         using a combination of {\tt diskbb}, {\tt power-law}, and {\tt Gaussian} models.}
\end{figure}

After finding the best-fitted continuum model, we included the $6.3-7.3$~keV energy band in our analysis, which is the main motivation of this 
work. We found that there is the presence of a triplet line emission in this energy band for the Resolve spectrum. Thus, to get the best fit, we added
three \textit{ga} lines, and our best-fitted model combination reads as {\tt tbabs*(diskbb + po + ga[3])} and achieves $\chi^2_{red} = 1.09$. The 
best-fitted unfolded spectrum is given in the upper panel of \autoref{fig:diskbb}.

The spectrum from the Resolve instrument showed the presence of multiple line emissions in the $6.3-7.3$~keV energy band. From the spectral 
fitting, we found that the $N_H$ was high with a value of $(16.7 \pm 0.1) \times 10^{22}$ cm$^{-2}$. The source is located in a very crowded region 
near the Galactic center. Due to its close proximity to the Galactic center, the MAXI satellite could not distinctly resolve the source. The inner 
disk temperature ($T_{in}$) is $0.62 \pm 0.04$ keV. The {\tt diskbb} model normalization, $Norm_{diskbb}$ obtained from the best fit, is $6339 \pm 
269$. We found that the photon index of the power-law was very high with a value of $\Gamma \sim 3.2 \pm 0.1$, suggesting that the source is in the 
soft state during this epoch of the outburst. We find that there was the presence of iron lines at $\sim$ 6.64, 6.67, and 6.99 keV, with widths $100 
\pm 30$, $35 \pm 3$, and $100 \pm 40$~eV, respectively. All the lines have a normalization value of the order of $10^{-4}$. These are possibly the 
multiple components of the Fe-XXV lines; however, they are subject to a detailed line profile study. In column 2 of \autoref{tab:diskbb}, the best
fitted model parameters are listed.

For the Xtend spectrum, as mentioned before, the line features are not resolved properly to identify them as multiple lines. Thus, we only added
one \textit{ga} component to the previously achieved best-fitted continuum for Xtend. Thus, our best-fitted model combination reads as {\tt tbabs*(diskbb
+ po + ga[2])} for which $\chi^2_{red} = 1.07$. The best-fitted unfolded spectrum is given in the lower panel of \autoref{fig:diskbb}.

For the Xtend, the line components were not resolved properly and looked like a broad Gaussian component. From this spectral fitting, we also
found that the $N_H$ was high with a value of $(17.9 \pm 0.1) \times 10^{22}$ cm$^{-2}$.  The inner disk temperature ($T_{in}$) is $0.60 \pm 0.05$ keV. 
The {\tt diskbb} model normalization, $Norm_{diskbb}$ obtained from the best-fit is $9812 \pm 578$. Here also, the high value of the photon index of the 
power-law $\Gamma \sim$ $3.3 \pm 0.1$ suggests the presence of the high soft state. We found the presence of an iron line at the line energy of $6.64 
\pm 0.01$~keV with a width of $161 \pm 17$~eV and a normalization of the order of $\sim 10^{-3}$. There was the presence of an excess flux at the low
energy of $\sim 2.36 \pm 0.02$ keV with a width of $70 \pm 2$~eV. However, the normalization of this Gaussian was $\sim 0.21 \pm 0.01$. This can be 
observed in the lower panel of \autoref{fig:diskbb}. As mentioned earlier also, we speculate it could be from the instrumental effect rather than astrophysical. 
In column 3 of \autoref{tab:diskbb}, the best-fitted model parameters are listed. We also estimated the line intensity ratio ($\frac{I_{\rm 6.9}}{I_{\rm 6.7}}$) 
of the $\sim$ 6.99 to $\sim$ 6.7 keV (adding both the lines at 6.64 and 6.67 keV). We found that it is $\sim$ 0.29, confirming the location of the source 
near the Galactic center \citep{2016ApJ...833..268N}.

\begin{table}[!h]
  \vskip 0.1cm
  \caption{\label{tab:diskbb}Values of the spectrally fitted parameters using {\tt tbabs}, {\tt diskbb}, {\tt power-law}, and multiple {\tt Gaussian} 
	   models. Column 1 represents the parameters of the phenomenological models used for spectral fitting. Columns 2 \& 3 represent the values of 
       those parameters for the Resolve and the Xtend instruments, respectively. The last value of flux is in the units of $10^{-9}~erg~cm^{-2}~s^{-1}$.}
  \centering
  \vskip 0.2cm
 \addtolength{\tabcolsep}{-1.5pt}
 \begin{tabular}{|c|c|c|}
 \hline
   (1)            &        (2)         &       (3)     \\
Parameters        &      Resolve       &      Xtend    \\
\hline            
  $N_H$           &  $16.7 \pm 0.1$    & $17.9 \pm 0.15$   \\
 $T_{in}$         &  $0.62 \pm 0.04$   & $0.60 \pm 0.05$   \\
$Norm_{diskbb}$   &  $6339 \pm 269$    & $9812 \pm 578$    \\
 $\Gamma$         &   $3.2 \pm 0.1$    & $3.3  \pm 0.1$    \\
  $Norm$          &  $1.21 \pm 0.27$   & $2.46 \pm 0.62$   \\
 $E_{Ga1}$        &  $6.64 \pm 0.11$   & $2.36 \pm 0.12$   \\
$\sigma_{Ga1}$    &  $0.10 \pm 0.04$   & $0.07 \pm 0.01$   \\
 $Norm_{Ga1}$     &  $2e-4 \pm 4e-5$   & $0.21 \pm 0.01$   \\
 $E_{Ga2}$        &  $6.67 \pm 0.17$   & $6.64 \pm 0.13$   \\
$\sigma_{Ga2}$    &  $0.03 \pm 0.01$   & $0.16 \pm 0.02$   \\
 $Norm_{Ga2}$     &  $3e-4 \pm 4e-5$   & $9e-4 \pm 8e-5$   \\
 $E_{Ga3}$        &  $6.99 \pm 0.02$   &        -          \\
$\sigma_{Ga3}$    &  $0.10 \pm 0.04$   &        -          \\
 $Norm_{Ga3}$     &  $1e-4 \pm 2e-5$   &        -          \\
 $\chi^2/DOF$     &   $6792/6235$      &   $1408/1321$     \\
   $Flux$         &  $1.05 \pm 0.01$   & $1.22 \pm 0.12 $  \\
\hline
 \end{tabular}
  \vskip 0.1cm
\end{table}

To further estimate the inner-disk radius i.e., the extent of the inner disk, we used the scaling relation, $Norm_{diskbb} = (r_{in}/D_{10})^2cos{\theta}$,
where $r_{in}$, $D_{10}$, and $\theta$ are the inner disk radius (in km), the distance of the source (in units of 10 kpc), and the inclination of the 
disk (in degrees) to the observer, respectively. However, the above-mentioned $r_{in}$ is subject to some uncertainty \citep{1995ApJ...445..780S, 
1998PASJ...50..667K}. The modified inner disk radius is given by $R_{in} \simeq \kappa^2 \xi r_{in}$, where $\kappa$ and $\xi$ are the hardening factor 
\citep{1998PASJ...50..667K} and inner boundary correction factor \citep{1995ApJ...445..780S}, respectively. According to them, these two correction 
factors have values of 1.7 and 0.41, respectively. Thus, the relation becomes $Norm_{diskbb} = (R_{in}/\kappa^2 \xi D_{10})^2cos{\theta}$. So, the 
correct innerdisk radius ($R_{in}$) is given as $R_{in} = (\kappa^2 \xi D_{10})\sqrt{(Norm_{diskbb}/cos{\theta})}$. According to \citet{2025ATel17063....1M}, 
we also assume a distance estimate of the source as $8$~kpc. Since the disk inclination of the source is not constrained, we assume values of $\theta 
\sim$ 20$^\circ$, 50$^\circ$, and 80$^\circ$ for low to high inclination. Given these values, the $R_{in}$ comes out to be $\sim$\{78, 94, and 184\}, 
and $\sim$\{97, 118, and 229\} km for Resolve and Xtend instruments, respectively. Considering the innermost stable circular orbit (ISCO) for the inner 
disk radius of a Schwarzschild black hole as a scale factor, the mass of this source varies from $8-25~M_\odot$. Due to the consideration of a broad 
range of inclination, the mass has also appeared in a broad range. Such estimates can provide preliminary information about the black hole mass as a 
parameter, but not the physical properties of the system.

\subsubsection{{\tt kerrbb} and {\tt power-law}}

\begin{figure}[!h]
\vskip 0.2cm
\centering
\vbox{
\includegraphics[width=8.5truecm,angle=0]{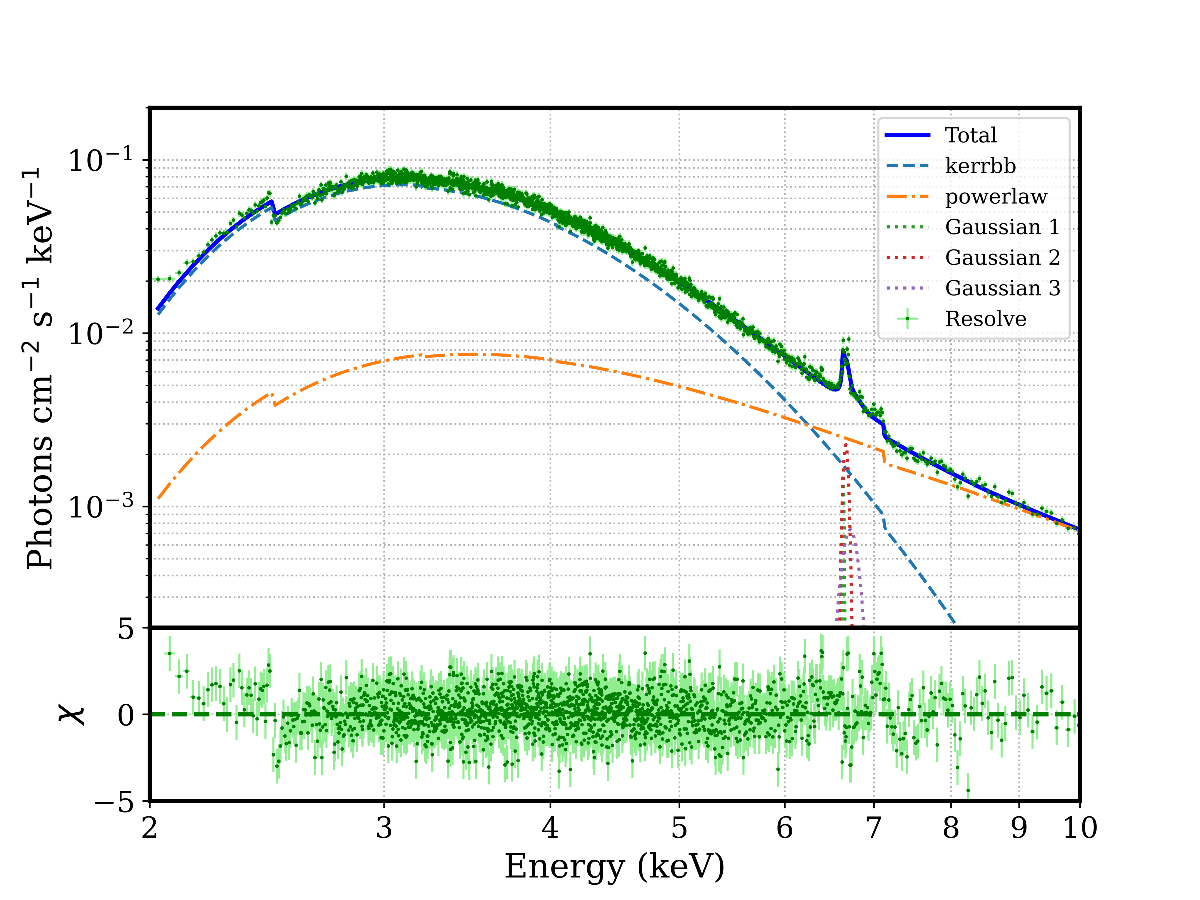}
\includegraphics[width=8.5truecm,angle=0]{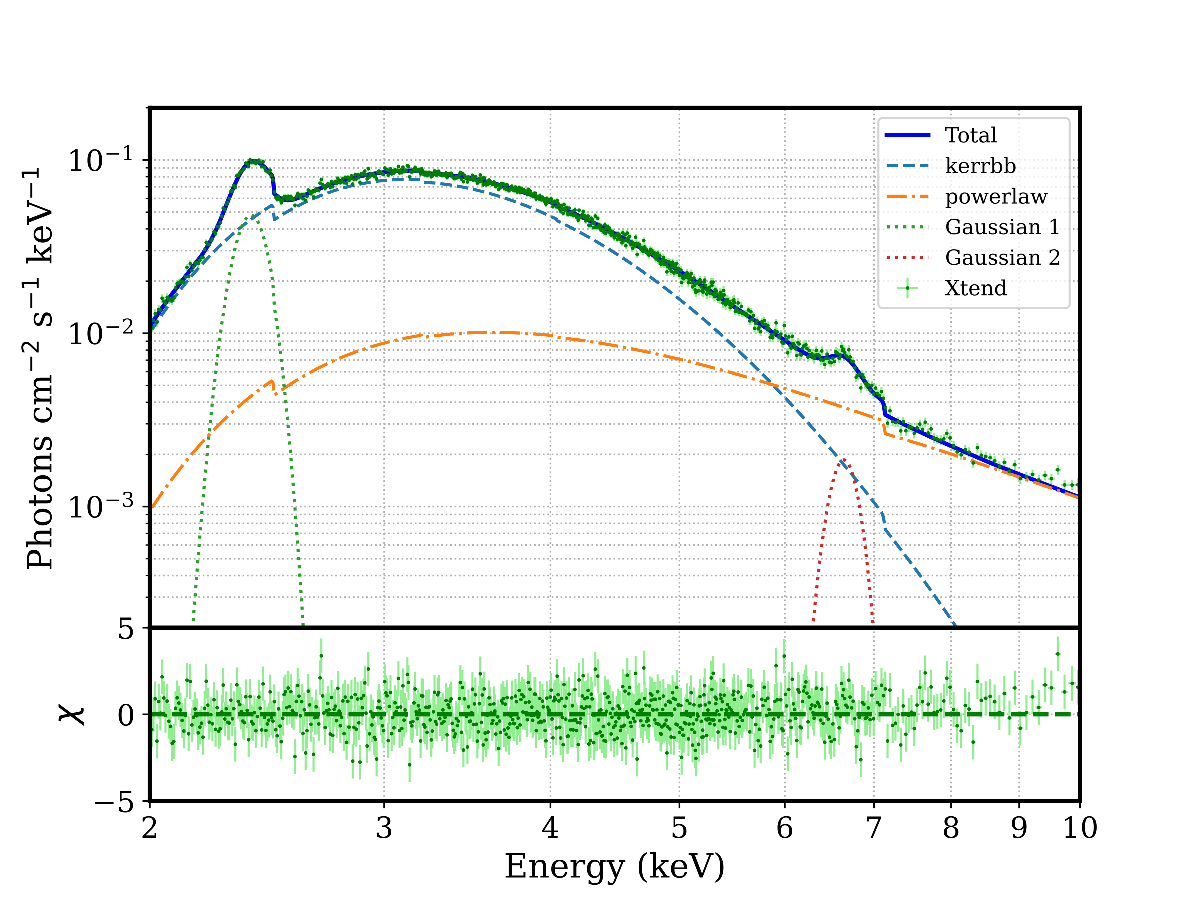}}
\caption{\label{fig:kerrbb} Model-fitted unfolded XRISM spectra for Resolve (top panel) and Xtend (bottom panel) instruments, respectively, by using a 
         combination of {\tt kerbb}, {\tt power-law}, and {\tt Gaussian} models.}
\end{figure}

We also performed the spectral analysis by using the {\tt \href{https://heasarc.gsfc.nasa.gov/xanadu/xspec/manual/node189.html}{kerrbb}} model
\citep{2005ApJS..157..335L} by replacing the {\tt diskbb} component. To perform the spectral fitting with this model, we have frozen the distance of the 
source to $8$~kpc \citep[][though not yet confirmed]{2025ATel17063....1M} and also frozen the hardening factor to 1.7. The use of {\tt kerrbb} model can 
give preliminary information about the intrinsic properties of the BH. Previously, the {\tt kerrbb} model has been used to constrain fundamental properties
of black hole systems like spin parameter during high soft state \citep{2018ApJ...867...86P}.

For Resolve, the best-fitted model combination reads as {\tt tbabs*(kerrbb + po + ga[3])} with $\chi^2_{red} \sim$ 1.09. The best-fitted unfolded
spectrum is shown in the upper panel of \autoref{fig:kerrbb}. For Xtend, the best-fitted model combination reads as {\tt tbabs*(kerrbb + po + ga[2])}, 
with a $\chi^2_{red} \sim$ 1.07. The best-fitted unfolded spectrum is shown in the lower panel of \autoref{fig:kerrbb}.

\begin{table}[h!]
  \vskip 0.1cm
  \caption{\label{tab:kerrbb}Values of the spectrally fitted parameters using {\tt tbabs}, {\tt kerrbb}, {\tt power-law}, and multiple {\tt Gaussian} 
           models. Column 1 represents the parameters of the models used for spectral fitting. Columns 2 \& 3 represent the values of those parameters 
           for the Resolve and the Xtend instruments, respectively. The last value of flux is in the units of $10^{-9}~erg~cm^{-2}~s^{-1}$.}
  \centering
  \vskip 0.2cm
 \addtolength{\tabcolsep}{-1.5pt}
 \begin{tabular}{|c|c|c|}
 \hline
   (1)            &        (2)         &       (3)     \\
Parameters        &      Resolve       &      Xtend    \\
\hline
  $N_H$           &  $17.0 \pm 0.2  $   & $18.1 \pm 0.7  $   \\ 
spin ($a$)        &  $0.68 \pm 0.02 $   & $0.65 \pm 0.02 $   \\
inclination ($i$) &  $19.8 \pm 0.3  $   & $23.4 \pm 0.3  $   \\
Mass ($M_{BH}$)   &  $8.8  \pm 0.3  $   & $9.6  \pm 0.5  $   \\
  $Mdd$           &  $0.86 \pm 0.03 $   & $0.87 \pm 0.04 $   \\
$Norm_{kerrbb}$   &  $1.7  \pm 0.2  $   & $2.27 \pm 0.3  $   \\
 $\Gamma$         &  $3.2  \pm 0.2  $   & $3.1  \pm 0.6  $   \\
  $Norm$          &  $1.25 \pm 0.13 $   & $1.62 \pm 0.25 $   \\
 $E_{Ga1}$        &  $6.64 \pm 0.14 $   & $2.36 \pm 0.11 $   \\
$\sigma_{Ga1}$    &  $0.01 \pm 0.00 $   & $0.07 \pm 0.01 $   \\
 $Norm_{Ga1}$     &  $8e-5 \pm 2e-5 $   & $0.22 \pm 0.02 $   \\
 $E_{Ga2}$        &  $6.67 \pm 0.32 $   & $6.64 \pm 0.14 $   \\
$\sigma_{Ga2}$    &  $0.03 \pm 0.01 $   & $0.16 \pm 0.02 $   \\
 $Norm_{Ga2}$     &  $2e-4 \pm 4e-5 $   & $9e-4 \pm 8e-5 $   \\
 $E_{Ga3}$        &  $6.72 \pm 0.03 $   &        -           \\
$\sigma_{Ga3}$    &  $0.10 \pm 0.01 $   &        -           \\
 $Norm_{Ga3}$     &  $2e-4 \pm 4e-5 $   &        -           \\
 $\chi^2/DOF$     &    $6801/6232$      &   $1406/1318$      \\
   $Flux$         &  $1.05 \pm 0.02 $   & $1.22 \pm 0.01 $   \\ 
\hline
 \end{tabular}
  \vskip 0.1cm
\end{table}

From this model fitting, we found that the disk has a low inclination of $\sim 19^\circ - 24^\circ$ and a moderate BH spin of $\sim 0.63-0.70$. This 
low inclination suggests that the inner disk might be at a distance of $\sim 75-100$ km from the central object, obtained from the previous relation. 
The {\tt kerrbb} model fit determines the mass to be $8.5-10.1$ M$_\odot$. The little disagreement between the two models in $M_{BH}$ estimation appears 
due to the choice of disk inclination. When performing the {\tt kerrbb} fitting, we kept the inclination as a free parameter and found a return value
in the 19-24 degrees, whereas for the {\tt diskbb} case, we assumed inclination values of 20, 50, and 80 degrees. If we consider the low inclination 
of $<25^\circ$ in {\tt diskbb} estimations, both models are providing a similar black hole mass. Since both models are phenomenological, to 
better constrain the parameter, data fitting using a more realistic model is required. Those estimations can be further verified and confirmed using 
the dynamical study of the source. Nonetheless, the present estimations further suggest that the object is a black hole and not a neutron star. 
Considering this mass, agrees that the inner boundary of the disk has moved inside the $6r_g$ distance from the object, which makes it suitable for 
iron line emission ($6r_g$ to $30r_g$). The results of this model fitting are given in \autoref{tab:kerrbb}.

\subsubsection{{\tt JeTCAF}}

In addition to these models, we have also implemented the physically motivated accretion-ejection based {\tt JeTCAF} model by replacing the {\tt PL} 
model to estimate the accretion-ejection parameters and the mass of the BH. Since JeTCAF has both hot (corona) and cold (disk) flow components, it can 
be fitted to any spectral states, whether a jet is present or not. Depending on the spectral state, model parameters can be constrained. As the source 
is in the soft state, that will be imprinted in the jet parameters as well, which can be further verified using this model. There are six parameters 
in this model, including the mass of the BH if it is not dynamically measured. The parameters are: (i) mass of the BH ($M_{BH}$ in solar mass $M_{\odot}$ 
units) (ii) cold Keplerian disk accretion rate ($\dot{m}_d$), (iii) hot sub-Keplerian halo accretion rate ($\dot{m}_h$) (both rates are in Eddington 
rate ${\dot{M}}_{Edd}$), (iv) size of the corona; shock location ($X_s$ in gravitational radius $r_g=GM_{BH}/c^2$), (v) density jump across the shock;  
compression ratio ($R = {\rho}_+/{\rho}_-$, where ${\rho}_+$ and ${\rho}_-$ represent the densities in the post-shock and pre-shock flows), and (vi) 
outflow collimation factor ($f_{col}$), the ratio of the solid angle subtended by the outflow to the inflow ($\Theta_{o}/\Theta_{i}$). Since the model 
has BH mass as a free parameter, one can successfully determine its value from the spectral modeling 
\citep[see][]{Debnathetal2014, MollaEtal2017ApJ...834...88M, 2024ApJ...975..257M}.

\begin{figure}[!h]
\vskip 0.2cm
\centering
\vbox{
\includegraphics[width=8.5truecm,angle=0]{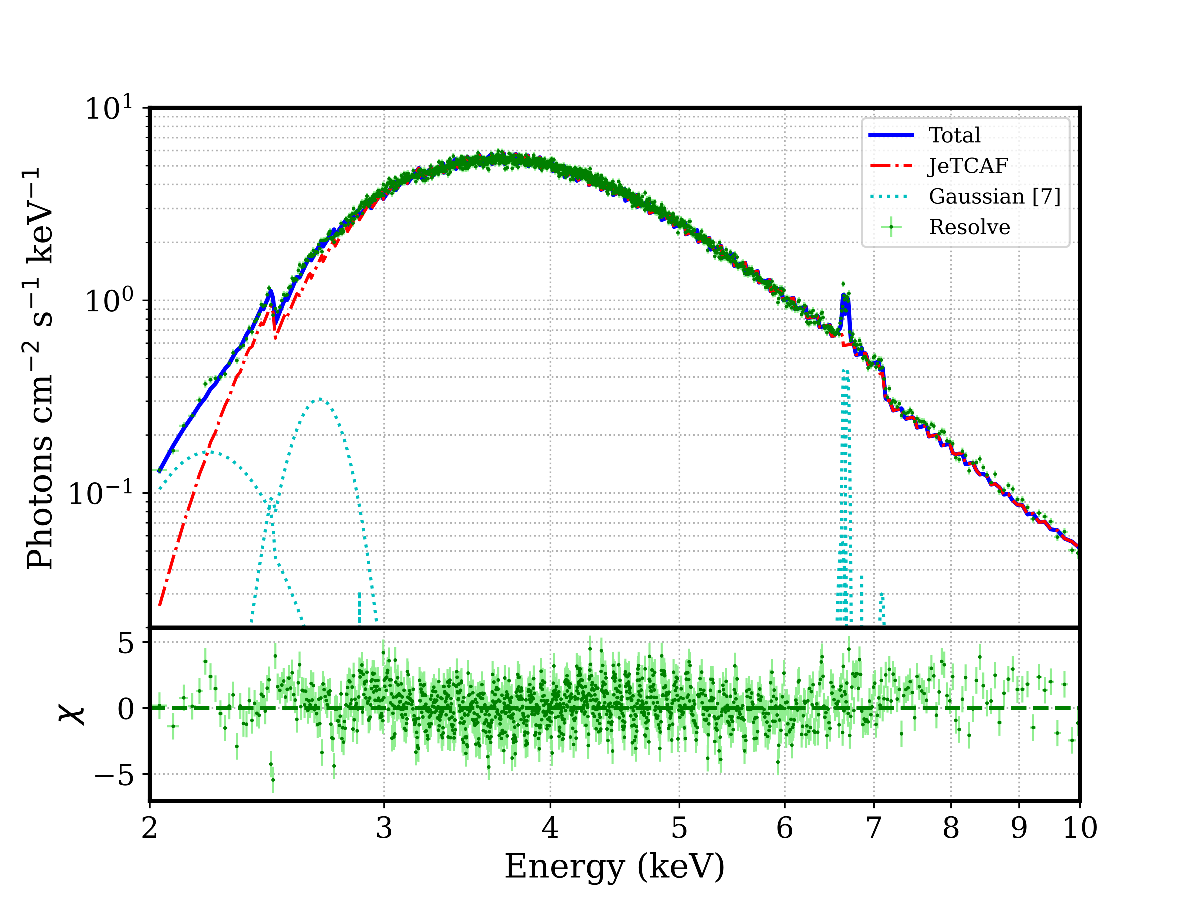}
\includegraphics[width=8.5truecm,angle=0]{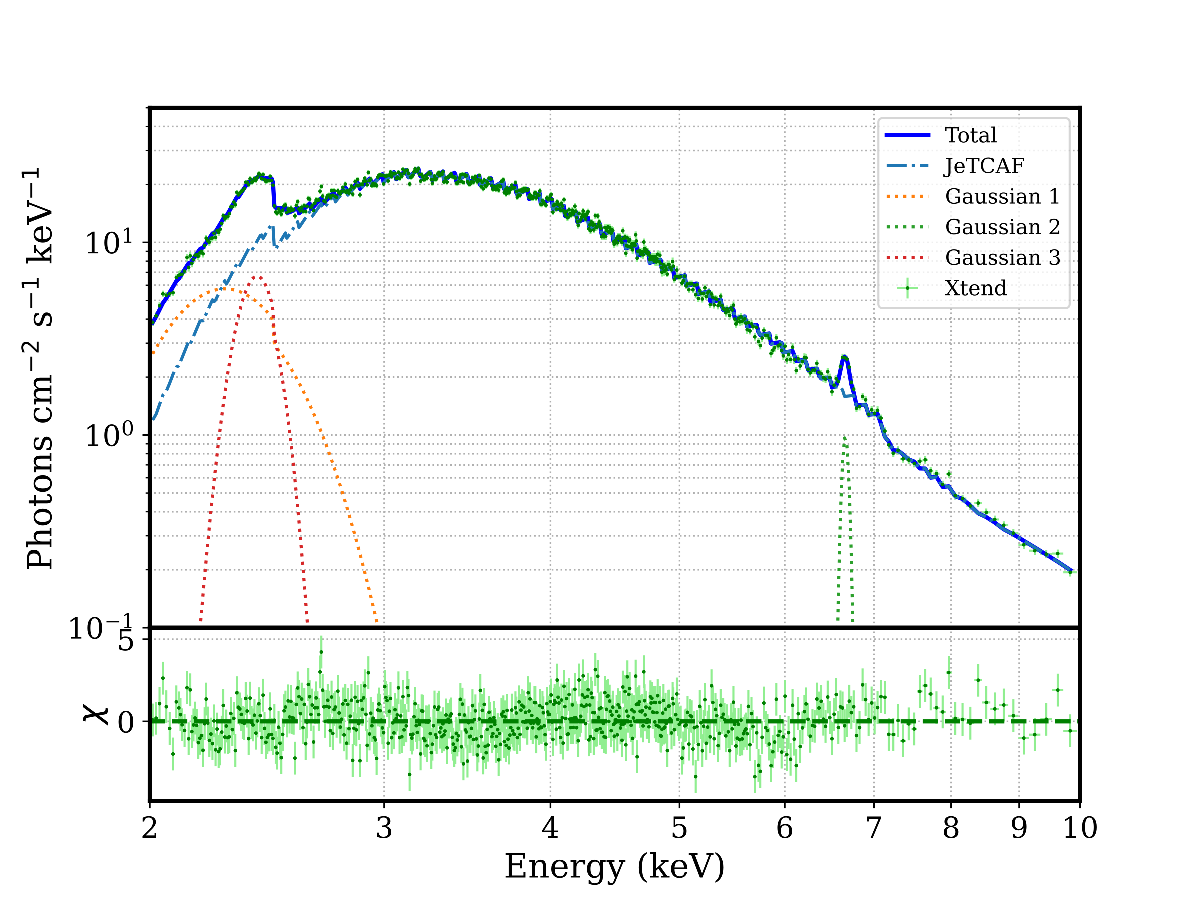}}
\caption{\label{fig:jetcaf}Best-fitted {\tt JeTCAF} model fitted spectra in the 2-10 keV energy band are shown. The top and bottom panels represent the 
         Resolve and Xtend instruments, respectively. The bottom windows of each panel show the residuals of the fits.} 
\end{figure}

The model combination reads in {\tt XSPEC} as: {\tt tbabs*(JeTCAF + ga[7])} for {\it Resolve}. The best-fitted unfolded spectrum could be seen in the 
upper panel of \autoref{fig:jetcaf}. Given the above model combinations, the {\it Resolve} spectrum is best-fitted with $\chi_{red}^2 = 1.3$ using {\tt
JeTCAF} model for the parameters $M_{BH} = 5.7\pm 0.8 M_\odot$, $\dot m_d=2.76\pm0.21$, $\dot m_h=0.25\pm0.06$, $X_s<12$, $R=6.52 \pm1.37$, and $f_{col}
= 0.12\pm0.03$. The high disk mass accretion compared to the hot sub-Keplerian component infers the soft spectral state. The Keplerian disk moved much 
closer to the BH $< 12 r_g$ (pegged at the lower bound of the parameter space), which is also associated with the effect of cooling due to the high disk 
accretion rate that cooled the hot corona. Such phenomena have also been observed for other LMBHs studied using the {\tt TCAF} model 
\citep{2014ApJ...786....4M, 2017ApJ...850...47M, 2021Ap&SS.366...63C, 2023ApJ...956...55C}. The high value of $R$ and low value of $f_{col}$ refer 
to the low mass outflow \citep[see][]{1999A&A...351..185C, 2021ApJ...920...41M} as expected in the soft state. Similar to other models, $N_H$ obtained 
from the fit is also high $(37.2 \pm 0.2) \times10^{22}$ cm$^{-2}$. Since there are multiple Fe K lines, we have frozen them to their peak energies at 6.62, 
6.64, 6.69, 6.85, and 7.1 keV and let their width vary. The best-fitted line width $\sigma$ for all Fe K lines is $<0.03$ keV.

For {\it Xtend} the best-fitted model combination reads as {\tt tbabs*(JeTCAF+ ga[3])}. The {\it Xtend} data equally fits well for a marginal change 
within error in the {\tt JeTCAF} model parameter as in {\it Resolve} with $\chi_{red}^2=1.2$. The $N_H$ value required for the {\it Xtend} data fitting 
is $(24.8 \pm 0.1) \times10^{22}$ cm$^{-2}$. The best-fitted model spectra are shown in \autoref{fig:jetcaf}.

The above model fittings help summarize the estimate of the mass of the BH. The {\tt diskbb} model fits determined the mass in the range $8-25 M_\odot$ 
for different values of the disk inclination angle, which is later better constrained by using the {\tt kerrbb} model fit. The estimated mass of the BH 
turns out to be $\approx 8-10 M_\odot$ from the above two models. From the {\tt JeTCAF} model fits, the mass of the BH comes out to be $\approx 5.7\pm
0.8 M_\odot$. Combining these three estimates, the most probable mass of the central compact object becomes $7.9\pm2.2 M_\odot$.

The estimation of $M_{BH}$ can be done using a more direct method from the dynamics of the stars orbiting the BH, which requires very high-resolution
optical telescopes. So far, only a handful of sources (less than $1/3$rd of total discovered BHCs; see \href{https://www.astro.puc.cl/BlackCAT/}{BlackCat}) 
are dynamically confirmed due to a lack of resolution \citep{2016A&A...587A..61C}. The mass of the BH is one of the fundamental quantities that is required 
to further study the system in greater detail. Therefore, we have estimated its value using the indirect method by modeling the X-ray spectra. Among the
different methods discussed above, the TCAF model or its variant (JeTCAF) has the potential to robustly measure the BH mass \citep{MollaEtal2017ApJ...834...88M,
2016MNRAS.460.3163M}, which has further produced consistent results in optical \citep[for MAXI J1659-152,][]{Corral-SantanaEtal2018MNRAS.475.1036C}. Additionally, 
a detailed and fully general relativistic treatment of mass estimation has also agreed with our mass estimation from spectral modeling \citep[e.g., for 
H 1743-322,][]{Tursunov2018PAN....81..279T}. In the same line, we believe that the mass estimated in this work for the present source may remain valid within 
error and can be verified further in the future using optical studies.

From all the modeling, we got very high values of the hydrogen column density ($N_H$). This may be because the source is situated in the Galactic 
center. The overly dense gas cloud \citep{2025PASJ...77L...1X} in the Galactic center may have blocked radiation, resulting in a high $N_H$.

\subsection{Photoionization Modeling ({\tt photemis})}

\begin{figure*}
    \centering
   \includegraphics[width=13cm,height=7cm]{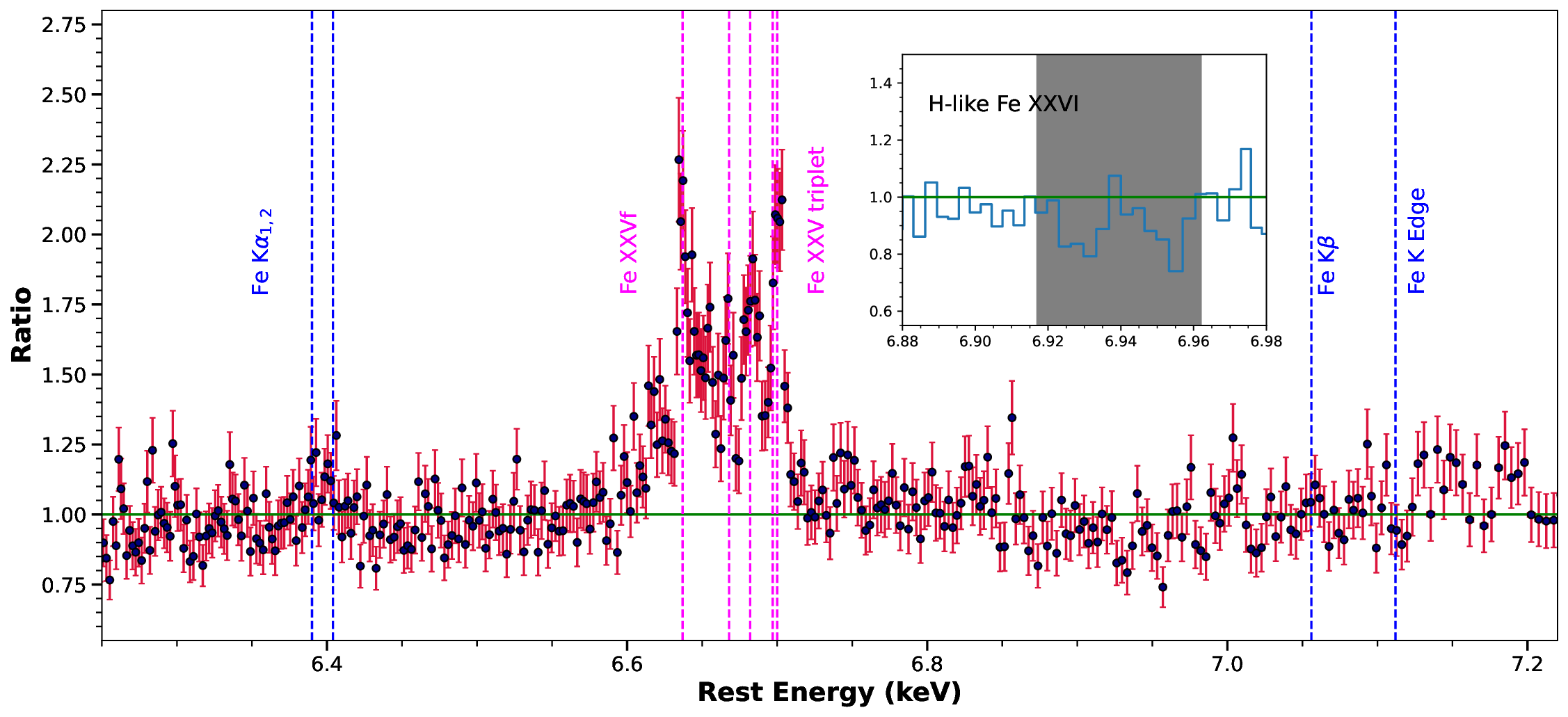}
   \includegraphics[width=14cm,height=7.5cm]{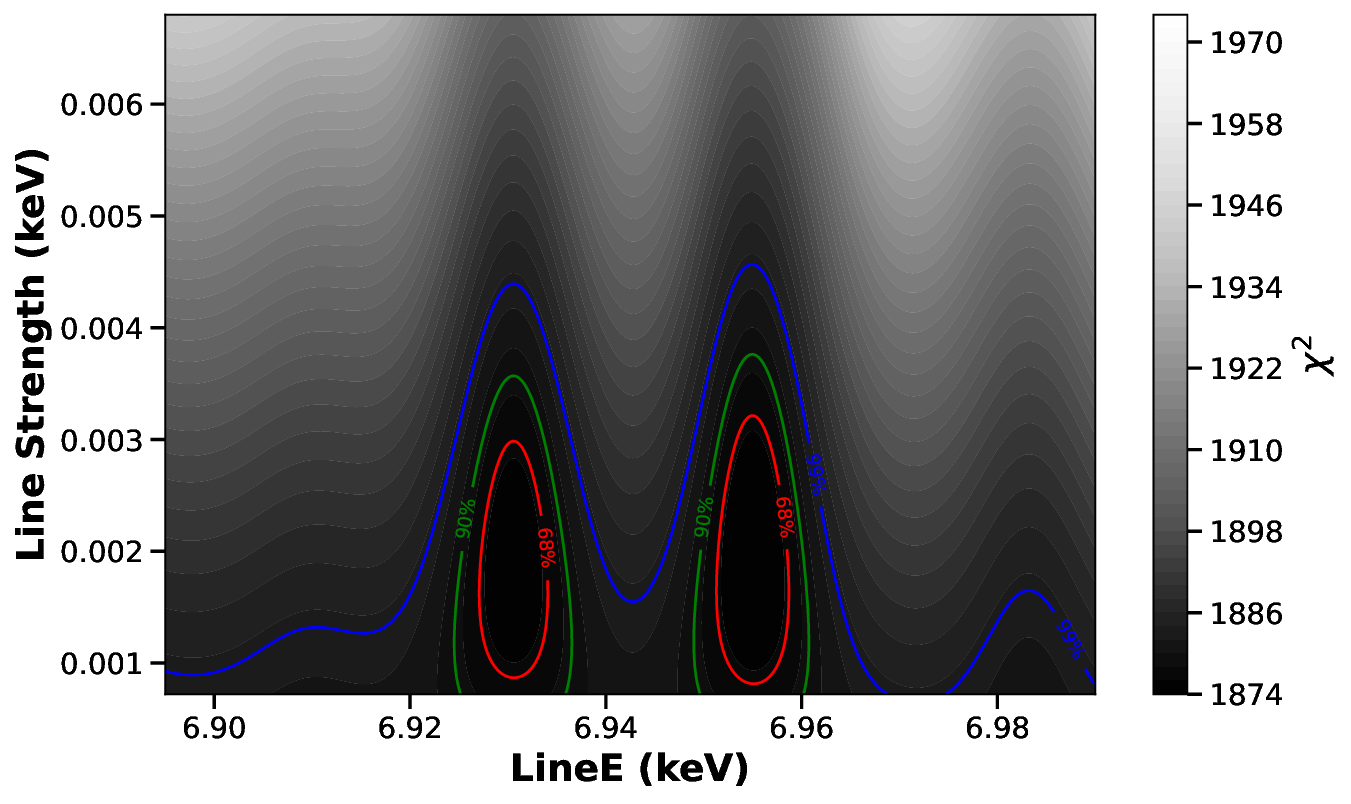}
    \caption{\textit{Top Panel}: Residuals from continuum subtracted 6-7 keV \textit{Resolve} spectrum showing wealth of complex Fe K lines marked using 
    vertical dashed lines. The He-like Fe XXV emission complex dominates the residuals, showing a mix of forbidden and recombination transitions. An inset plot 
    shows a zoomed-in view of 6.9 keV, capturing weak H-like Fe XXVI absorption line features. \textit{Bottom Panel}: \texttt{steppar} results for absorption line
    energy and normalization after scanning a narrow Gaussian absorption to the continuum model. The red, green, and blue contours correspond to 68\%, 90\%, and 
    99\% confidence levels.}
    \label{fig:lines}
\end{figure*}

\begin{figure*}
    \centering
   \includegraphics[width=13cm,height=8cm]{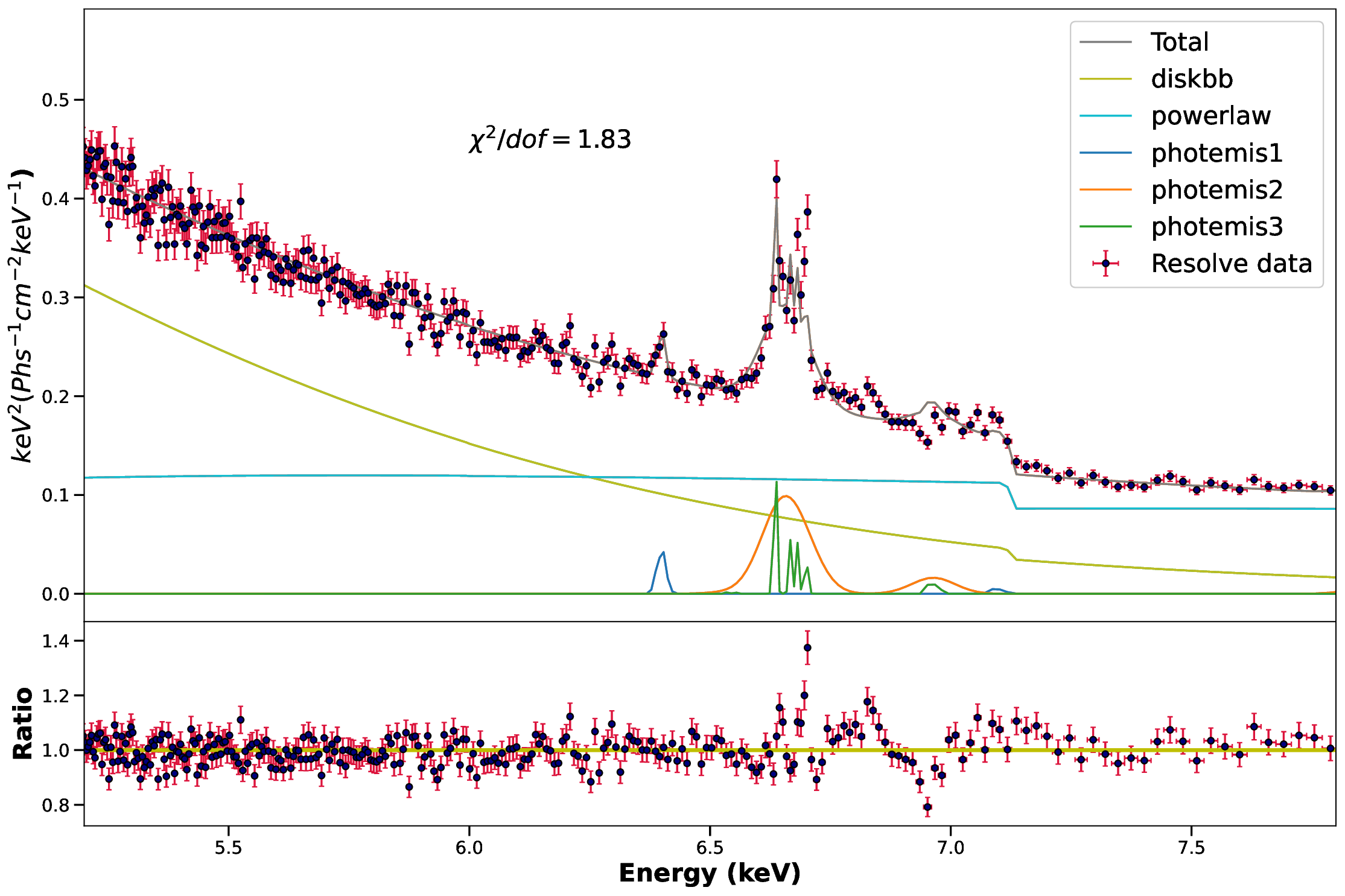}
    \caption{Best fitted \texttt{photemis} model fitted to the 5-8 keV \textit{Resolve} spectrum is shown. The top and bottom panels represent the 
            decomposition of model components and residuals, respectively.}
    \label{fig:photemis}
\end{figure*}

The source MAXI J1744-294 showed complex Fe emission features between 6 - 7 keV in the \textit{Resolve} spectrum. The left panel of \autoref{fig:lines} 
shows the continuum subtracted $6-7$ keV \textit{Resolve} spectrum. We clearly detected a weak Fe K alpha doublet (K$\alpha_{\rm 1,2}$) along with Fe K$\beta$ 
and K$\alpha$-edge. Strong He$\alpha$ like complex Fe XXV (forbidden and triplet) emission lines between 6.6 - 6.7 keV were detected. As shown in the inset, weak
absorption features corresponding to Fe XXVI were detected. We added a narrow absorption component \texttt{gabs} to continuum model and scanned the 6.8 - 7 keV 
range using \texttt{steppar} command. While the width was frozen to 5 eV, the line energy centroid and line strength were varied for 1000 x 1000 iterations. The 
right panel of \autoref{fig:lines} shows a weak, but significant detection of Fe XXVI absorption feature, which is red-shifted to $\sim$ 6.94 keV from rest frame 
by $z$= -0.0043. So far, we have used phenomenological multiple Gaussian components to fit different Fe K line transitions. However, to produce the line profiles, 
a realistic emission model is favorable. Therefore, we adopted the publicly available \texttt{photemis} model, derived from the \texttt{XSTAR} photoionization 
code \citep{2001ApJS..133..221K}. It consists of analytically computed spectra produced by recombined and collisionally ionized plasma.  

While fitting the \textit{Resolve} data between 5 - 8 keV, we kept the line of sight N$_{\rm H}$ and T$_{\rm in}$ fixed to prior obtained values of 16.7 
$\times$ 10$^{22}$ cm$^{-2}$ and 0.62 keV, as they cannot be constrained in the narrow band spectrum. The spectra was re-binned to a minimum of 500 counts 
per bin. We used the standard pre-calculated ion population of \texttt{xstar}, setting all abundances to zero except for iron, which was frozen to 1. The 
remaining free parameters were ionization parameter ($\log \xi$), turbulent velocity (v$_{\rm turb}$) and normalization (Norm$_{\rm pht}$). We found that 
a combination of three \texttt{photemis} components was able to satisfactorily describe the spectral line features between 6-7 keV. The best-fitted model 
decomposition along with residuals is shown in \autoref{fig:photemis}. The best fit values are presented in \autoref{tab:photemis_fit}. 

The photoionized emission was modeled by a low ionization component ($\log \xi \sim$ 1.56) that accounts for Fe K$\alpha$ emission around 6.40 keV and two 
highly ionized components ($\log \xi \sim$ 3) that wholly reproduce the Fe XXV emission complex. We tied the turbulent velocity of two narrow components
(\texttt{photemis} \#1 and \#3) and while keeping the broad component (\texttt{photemis} \#2) free during the fit. Our best fitted results 
indicated that the Fe XXV complex originates from two comparably ionized plasmas of $\log \xi$= 3.15 and 3.17, however, showing very distinct velocity widths. 
The broad component with v$_{\rm turb}$= 2513 km s$^{-1}$ is most likely produced from extremely hot gas located close to the inner edge of the accretion disk. 
The lines are broadened under the effect of Keplerian motion of the accretion disk as it extends close to the black hole during the soft spectral state. 
On the other hand, the narrow component with v$_{\rm turb}$ = 153 km s$^{-1}$ is probably the scattered emission from nearby gas that has been strongly 
photoionized by hard X-ray radiation. Another narrow, low ionized component ($\log \xi \sim$ 1.6) fits the weak Fe K$\alpha$ line, which originates 
further away from the accreting source, in a much cooler and less dense region.   

Therefore, our analysis of multiple Fe K emission lines in the soft state in this source reveals contributions from neutral, He-like, and H-like
iron. This is unusual compared to the traditional picture of a single, uniform reflection component for XRBs, indicating instead a stratified reprocessing 
medium spanning the accretion disk and possibly disk winds in the soft state. The distinct line components allow separate constraints on ionization rates, 
velocity broadening, and possibly the geometry of the medium (subject to detailed modeling), providing further detail on the ionization gradient of the 
reflector. Moreover, these results suggest a possible coupling between inflow and outflow in the soft state of the accreting systems, making the source more
interesting to further study with possible potential applications to other transients.

Since this source is located close to the galactic center, the emission line complex is expected to be contaminated by X-ray diffuse emission from stellar
populations in the Galactic bulge and unresolved, accreting magnetic cataclysmic variables (mCVs) \citep{Rev2009Natur.458.1142R,anas2023A&A...671A..55A}. 
This could potentially explain the residuals around $\sim$ 6.7 and 6.8 keV. The persistent narrow emission feature around 6.85 keV does not correspond to 
any recognized atomic transition. We speculate that this could be due to a redshifted Fe XXVI transition originating from ionized, outflowing winds. We 
also noticed clear, broad residuals around 7 keV as well, which, most likely, is due to the blended emission from Fe XXVI Ly$\alpha$ and Fe K$\beta$ lines, 
requiring sophisticated handling of photoionization models, which is beyond the scope of this work. 

\begin{table}[!h]
\scriptsize
\centering
\caption{Best-fit parameters of the photoionization model}
\hspace{-1cm}
    \begin{tabular}{|c|c|c|c|}
    \hline
         Component & $\log \xi$ &v$_{\rm turb}$ (km s$^{-1}$) &Norm$_{\rm pht}$  \\
         \hline
         \texttt{photemis} \#1 &1.55 $\pm$ 0.05 & 153.29 $\pm$ 54.56 & 1561.36 $\pm$ 293.49\\
         
         \texttt{photemis} \#2  &3.17 $\pm$ 0.15 &2513.30 $\pm$ 206.48 &522.95 $\pm$ 38.24\\
         \texttt{photemis} \#3 &3.15 $\pm$ 0.05 &= \#1 &94.13 $\pm$ 18.48 \\
         \hline
    \end{tabular}
    \label{tab:photemis_fit}
\end{table}

Presence of a significant and strong Fe absorption feature in the X-ray spectra is often interpreted as characteristic of high inclination 
X-ray binaries \citep{2003A&A...407.1079B, 2004A&A...418.1061B}. Despite the high spectral resolution and high signal-to-noise ratio in 6-7 keV, 
the detection of a weak absorption feature (\autoref{fig:lines} and \autoref{fig:photemis}) supports the low inclination, non-dipping nature of 
the source. This is further supported by our estimation of low inclination (19-24 degrees) from continuum disk modeling using {\tt kerrbb} (see 
\autoref{tab:kerrbb}). Interestingly, the detection of a mild redshifting in this absorption feature could be a signature of failed winds, similar 
to that observed in BH transient- 4U 1630-472 \citep{2025ApJ...988L..28M}. At a few \% of Eddington luminosity, the source MAXI J1744-294 could 
be hosting such winds that were launched vertically from the disk and became bound to the system. Follow-up monitoring, especially during the 
quiescent and outburst phases, can provide more insights into the gaseous kinematics of this system.

\section{Conclusions}

We have studied the spectral properties of the BHC MAXI J1744-294 after its very first detection in 2025. Using publicly available archived XRISM data
on March 03, 2025 (MJD 60737), we have studied the spectrum of the source using both the instruments of this satellite, namely, Resolve and Xtend. The
spectacular spectral resolution of the XRISM satellite, especially the Resolve, helps us find the existence of multiple iron line profiles in the spectra.
In Xtend, we also found the existence of an iron line. To take care of the continuum emission, we have used various combinations of phenomenological
and physical models ({\tt diskbb}, {\tt kerbb}, {\tt power-law}, {\tt JeTCAF}, and {\tt photemis}). To take care of the line emissions, we have used 
multiple lines in the form of the {\tt Gaussian} model with continuum models. Along with these, we have used an interstellar absorption model, 
{\tt tbabs}. We have achieved the best fits with these models. We have also implemented the photoionization modeling in the same spectra using the
\texttt{photemis} analytic models implemented in \texttt{XSPEC}. This is one of the detailed studies of both continuum and line profiles of 
a low-mass X-ray binary in recent times. From our model fitted results, we conclude that:

\begin{itemize}

\item The system is situated in a crowded region close to the Galactic center, supported by the ratios of the Iron lines, resulting in a high
hydrogen column density. The data could be well explained by models with disk inclination of $i \sim 19^\circ-24^\circ$, black hole spin of $a \sim 
0.63-0.70$, and a mass of $M_\odot \sim 7.9 \pm 2.2$~M$_\odot$.

\item From the values of the photon index, inner-disk temperature, inner-disk radius, and accretion rates, we conclude that the source was in 
the soft spectral state during this time of outburst activity.

\item High-resolution XRISM spectroscopy has revealed, for the first time, complex iron line features in this source, corresponding to distinct 
components of Fe XXV emission along with Fe XXVI absorption features. 

\item These Fe XXV line complexes arise from two highly ionized plasmas ($\log \xi \sim$ 3) with distinct turbulent velocities—one broad (v$_{\rm turb} 
\approx$ 2513 km s$^{-1}$) from hot gas at the inner accretion disk, and one narrow (v$_{\rm turb} \approx$ 153 km s$^{-1}$) scattered by nearby photoionized 
gas. A separate low-ionization component ($\log \xi \sim$ 1.6) accounts for the weak, narrow Fe K$\alpha$ fluorescent line from cooler, distant regions. 
These results offer new insight into the reprocessing of continuum in stratified media, either in the accretion disk or winds, or both, for XRBs 
in the soft state.

\end{itemize}

\section{Data Availability}

This work has made use of publicly archived data from the XRISM satellite, which is a mission of the Japan Aerospace Exploration Agency in partnership
with NASA and ESA. This work has also made use of software from the HEASARC, which is developed and monitored by the Astrophysics Science Division at
NASA/GSFC and the High Energy Astrophysics Division of the Smithsonian Astrophysical Observatory.

\section{Acknowledgements}

We thank the referee for making valuable comments and suggestions. We thank Michael Loewenstein of the University of Maryland, College Park, USA, 
for his assistance during the data reduction. K. C., C. B. S., B. K., \& X.-W. L. acknowledge support from the ``Science \& Technology Champion Project'' 
(202005AB160002) and from two ``Team Projects'' -- the ``Top Team'' (202305AT350002) and the ``Innovation Team'' (202105AE160021), all funded by the 
``Yunnan Revitalization Talent Support Program''. They also acknowledge the support from the "Key Laboratory of Survey Science of Yunnan Province" with 
project No. 202449CE340002. S. M. acknowledges the Ramanujan Fellowship (\# RJF/2020/000113) by SERB/ANRF-DST, Govt. of India, for this research. B.P. 
acknowledges collective support from Narodowe Centrum Nauki (NCN) grants 2021/41/B/ST9/04110 and 2018/31/G/ST9/03224 for this research. C. B. S. is also 
supported by the National Natural Science Foundation of China under grant no. 12073021. B. K. also acknowledges the support from the `Special Project for 
High-End Foreign Experts, Xingdian Funding from Yunnan Province, and the National Key Research and Development Program of China (2024YFA1611603). H.-K. C. 
acknowledges support from the NSTC project (grant No. 114-2112-M-007-042) of NTHU.

\software{
{\tt HEASoft} (version: 6.35.1), 
{\tt XSPEC} (version: 12.15.0) \citep{1996ASPC..101...17A}, 
Matplotlib (version: 3.6.1) \citep{hunter2007matplotlib},
{\tt XSTAR} (version: 2.58) \citep{2001ApJS..133..221K},
XRISM Pipelines with {\tt caldb} (version: 20250315 for Resolve; version: 20241115 for Xtend and GEN)
}

\bibliography{ref}{}
\bibliographystyle{aasjournalv7}

\end{document}